\documentclass[aps,prl,showpacs,showkeys,groupedaddress,preprint]{revtex4}

\usepackage[colorlinks]{hyperref}
\usepackage{color}
\usepackage{latexsym}
\usepackage{bm}
\usepackage{amsmath}
\usepackage{amssymb}
\usepackage{amsfonts}

\newcommand{\bracr}[1]   {\left({#1}\right)}

\newcommand{\bracp}[1]   {\left\lvert{#1}\right\rvert}

\newcommand{\lcond}[2]   {\left.{#1}\,\right|{#2}}

\DeclareMathOperator{\varop}	{var}
\DeclareMathOperator{\covop}	{cov}

\DeclareMathOperator{\traceop}	{trace}
\DeclareMathOperator{\transop}	{^\intercal}
\DeclareMathOperator{\Covop}	{\mathbf\Sigma}

\DeclareMathOperator{\TE}		{\mathcal T}
\DeclareMathOperator{\GC}		{\mathcal F}
\DeclareMathOperator{\EGC}		{\widehat\GC}

\newcommand{\var}[1]			{\varop\!\bracr{#1}}

\newcommand{\cov}[2]			{\covop\!\bracr{{#1},{#2}}}
\newcommand{\covs}[1]			{\covop\!\bracr{{#1}}}

\newcommand{\ccovs}[2]			{\covs{\lcond{#1}{#2}}}

\newcommand{\entro}[1]			{H\!\bracr{#1}}
\newcommand{\centro}[2]			{\entro{\lcond{#1}{#2}}}

\newcommand{\dett}[1]			{\bracp{#1}}
\newcommand{\trace}[1]			{\traceop\!\bracr{#1}}
\newcommand{\trans}[1]			{{#1}\!\transop}
\newcommand{\lnt}[1]			{\ln\!\bracr{#1}}

\newcommand{\Covs}[1]			{\Covop\!\bracr{#1}}
\newcommand{\Cov}[2]			{\Covop\!\bracr{{#1},{#2}}}
\newcommand{\cCov}[2]			{\Covop\!\bracr{\lcond{#1}{#2}}}

\newcommand{\cgc}[3]			{\GC_{\lcond{{#1}\rightarrow{#2}}{#3}}}
\newcommand{\ecgc}[3]			{\EGC_{\lcond{{#1}\rightarrow{#2}}{#3}}}
\newcommand{\cxe}[3]			{\TE_{\lcond{{#1}\rightarrow{#2}}{#3}}}

\newcommand{\bX}				{\bm X}
\newcommand{\bY}				{\bm Y}
\newcommand{\bZ}				{\bm Z}
\newcommand{\blX}				{\bX^{\!-}}
\newcommand{\blY}				{\bY^{\!-}}
\newcommand{\blZ}				{\bZ^{\!-}}
\newcommand{\bx}				{\bm x}
\newcommand{\by}				{\bm y}

\newcommand{\eps}				{\varepsilon}
\newcommand{\beps}				{\bm\eps}
\newcommand{\balpha}			{\bm\alpha}
\newcommand{\mc}   				{\oplus}

\newcommand{\shalf}[0]			{\tfrac 1 2}

\newcommand{\eeqref}[1]			{eq.~\eqref{#1}}

\newcommand{\eg}[0]				{e.g.}
\newcommand{\ie}[0]				{i.e.}

\newcommand{\cf}[0]				{\textit{cf.}}

\begin{document}

\title{Granger causality and transfer entropy are equivalent for Gaussian variables}

\author{Lionel Barnett}
\email{L.C.Barnett@sussex.ac.uk}

\affiliation{
	Centre for Computational Neuroscience and Robotics \\
	School of Informatics \\
	University of Sussex \\
	Brighton BN1 9QJ, UK
}

\author{Adam B. Barrett}
\email{abb22@sussex.ac.uk}

\author{Anil K. Seth}
\email{A.K.Seth@sussex.ac.uk}

\affiliation{
	Sackler Centre for Consciousness Science \\
	School of Informatics \\
	University of Sussex \\
	Brighton BN1 9QJ, UK
}

\date{\today}

\begin{abstract}

\emph{Granger causality} is a statistical notion of causal influence based on prediction via vector autoregression.  Developed originally in the field of econometrics, it has since found application in a broader arena, particularly in neuroscience. More recently \emph{transfer entropy}, an information-theoretic measure of time-directed information transfer between jointly dependent processes, has gained traction in a similarly wide field. While it has been recognized that the two concepts must be related, the exact relationship has until now not been formally described. Here we show that for Gaussian variables, Granger causality and transfer entropy are entirely equivalent, thus bridging autoregressive and information-theoretic approaches to data-driven causal inference.

\end{abstract}



\maketitle

The problem of inferring causal interactions from data has challenged scientists and philosophers for centuries \cite{Pearl99}.  One approach that has become increasingly popular over recent years was introduced originally by Wiener \cite{Wiener:1956}, and formalized in terms of linear autoregression by Granger \cite{granger69}. According to Wiener-Granger causality (G-causality), given sets of inter-dependent variables $\bX$ and $\bY$, it is said that ``$\bY$ G-causes $\bX$'' if, in an appropriate statistical sense, $\bY$ assists in predicting the future of $\bX$ beyond the degree to which $\bX$ already predicts its own future. Importantly, identification of a G-causality interaction is \emph{not} identical to identifying a physically instantiated causal interaction in a system. Although the two descriptions are intimately related \cite{SethEdelman07,cadotte:2008}, physically instantiated causal structure can only be unambiguously identified by perturbing a system and observing the consequences \cite{Pearl99}.  Nonetheless, G-causality is pragmatic, well-defined, and has delivered many insights into the functional connectivity of systems in a variety of fields, particularly in neuroscience \cite{DingEtal06}.

The information-theoretic notion of \emph{transfer entropy} was formulated by Schreiber \cite{schreiber00} as a measure of directed (time-asymmetric) information transfer between joint processes. In contrast to G-causality, transfer entropy is framed not in terms of prediction but in terms of \emph{resolution of uncertainty}. One can say that ``the transfer entropy from $\bY$ to $\bX$'' is the degree to which $\bY$ disambiguates the future of $\bX$ beyond the degree to which $\bX$ already disambiguates its own future. There is therefore an attractive symmetry between the notions (``predicts'' $\leftrightarrow$ ``disambiguates'') which has been noted previously (see \eg\ \cite{Palus01}) but never explicitly specified. In this Letter we show that under Gaussian assumptions they are in fact entirely equivalent. Our results therefore provide a framework for inferring causality which unifies information-theoretic and autoregressive approaches.
\newline

We use a standard mathematical vector/matrix notation in which bold type generally denotes vector quantities and upper-case type denotes matrices or random variables, according to context. All vectors are considered to be \emph{row} vectors. The symbol `$\transop$' denotes the transpose operator and `$\mc$' denotes \emph{concatenation} of vectors, so that for $\bx = (x_1,\ldots,x_n)$ and $\by = (y_1,\ldots,y_m)$, $\bx\mc\by$ is the $1 \times (n+m)$ vector $(x_1,\ldots,x_n,y_1,\ldots,y_m)$.

Given jointly distributed multivariate random variables (\ie\ random vectors) $\bX,\bY$, we denote by $\Covs\bX$ the $n \times n$ matrix of covariances $\cov{X_i}{X_j}$ and by $\Cov\bX\bY$ the $n \times m$ matrix of cross-covariances $\cov{X_i}{Y_\alpha}$. We then use $\cCov\bX\bY$ to denote the $n \times n$ matrix
\begin{equation}
    \cCov\bX\bY \equiv \Covs\bX - \Cov\bX\bY \Covs\bY^{-1} \trans{\Cov\bX\bY} \label{eq:ccxy}
\end{equation}
defined when $\Covs\bY$ is invertible. $\cCov\bX\bY$ appears as the covariance matrix of the residuals of a linear regression of $\bX$ on $\bY$ [\cf\ \eeqref{eq:rescov} below]; thus, by analogy with \emph{partial correlation} \cite{Kendall79} we term $\cCov\bX\bY$ the \emph{partial covariance} \footnote{This is to be distinguished from the \emph{conditional covariance}, which will in general be a random variable - although later we note that for \emph{Gaussian} variables the notions coincide.} of $\bX$ given $\bY$.

Suppose we have a multivariate stochastic process $\bX_t$ in discrete time \footnote{While our analysis may be extended to \emph{continuous} time we focus here on the discrete time case.} (\ie\ the random variables $X_{ti}$ are jointly distributed). We use the notation $\bX^{(p)}_t \equiv \bX_t\mc\bX_{t-1}\mc\ldots\mc\bX_{t-p+1}$ to denote $\bX$ itself, along with $p-1$ \emph{lags}, so that $\bX^{(p)}_t$ is a $1 \times pn$ random vector for each $t$. Given the lag $p$, we use the shorthand notation $\blX_t \equiv \bX^{(p)}_{t-1}$ for the lagged variable.
\newline

Let $\bX, \bY$ be jointly distributed random vectors and consider the linear regression
\begin{equation}
    \bX = \balpha + \bY \cdot A + \beps \label{eq:linreg}
\end{equation}
where the $m \times n$ matrix $A$ comprises the regression coefficients, $\balpha = (\alpha_1,\ldots,\alpha_n)$ are the constant terms and the random vector $\beps = (\eps_1,\ldots,\eps_n)$ comprises the residuals. The mean squared error (MSE) may then be written in terms of the covariance matrix of the residuals as $E^2 = \trace{\Covs\beps}$. $E^2$ is just the sum of the variances of the $\eps_i$, sometimes known as the \emph{total variance}. Performing an Ordinary Least Squares (OLS) to find the coefficients $A$ that minimize $E^2$ yields [assuming $\Covs\bY$ invertible] $A = \Covs\bY^{-1} \trans{\Cov\bX\bY}$ and we find that for the least squares fit the covariance matrix of the residuals is given by
\begin{equation}
    \Covs\beps = \cCov\bX\bY \label{eq:rescov}
\end{equation}
with $\cCov\bX\bY$ the partial covariance as defined by \eqref{eq:ccxy}. We note that the same coefficients $A$  which minimize the total variance $E^2$ also minimize the \emph{generalized variance} $\dett{\Covs{\beps}}$ \cite{Wilks32}, where $\dett{\,\cdot\,}$ denotes the determinant (this procedure is sometimes referred to as ``Least Generalized Variance''; see \eg\ \cite{Davidson00}).

If the residuals $\beps$ can be taken to be uncorrelated with the regressors $\bY$ in \eqref{eq:linreg}---as would be the case, for instance, for a multivariate autoregressive (MVAR) model---the residual covariance matrix can be derived directly from \eqref{eq:linreg}. Taking the covariance of both sides of \eqref{eq:linreg} yields
\begin{equation}
    \Covs\bX = \trans A \Covs\bY A + \Covs\beps \label{eq:sigx}
\end{equation}
Since the residuals and regressors are uncorrelated, we also have
\begin{eqnarray}
    0
	&=& \Cov\bY\beps \nonumber \\
	&=& \Cov\bY{\bX-\balpha-\bY \cdot A} \nonumber \\
	&=& \trans{\Cov\bX\bY} - \Covs\bY A \label{eq:yweq}
\end{eqnarray}
Solving \eqref{eq:yweq} for $A$ and substituting in \eqref{eq:sigx} we recover \eeqref{eq:rescov} for $\Covs\beps$. We note that eqs.~\eqref{eq:sigx} and \eqref{eq:yweq} are essentially \emph{Yule-Walker} equations \cite{DingEtal06} for the regression \eqref{eq:linreg}.

Suppose now we have three jointly distributed, stationary \footnote{The analysis carries through for the non-stationary case, but for simplicity we assume here that all processes are stationary.} multivariate stochastic processes $\bX_t, \bY_t, \bZ_t$ (``variables'' for brevity). Consider the regression models:
\begin{eqnarray}
    \bX_t &=& \balpha_t  + \bracr{\bX^{(p)}_{t-1}\mc\bZ^{(r)}_{t-1}} \cdot A  + \beps_t\label{eq:reg1} \\
    \bX_t &=& \balpha'_t + \bracr{\bX^{(p)}_{t-1}\mc\bY^{(q)}_{t-1}\mc\bZ^{(r)}_{t-1}} \cdot A' + \beps'_t \label{eq:reg2}
\end{eqnarray}
so that the ``predictee'' variable $\bX$ is regressed firstly on the previous $p$ lags of itself plus $r$ lags of the conditioning variable $\bZ$ and secondly, in addition, on $q$ lags of the ``predictor'' variable $\bY$ \footnote{Eqs. \eqref{eq:reg1} and \eqref{eq:reg2} are not to be interpreted as specifying actual MVAR processes; indeed, as such they could not be consistent due to the $\bY^{(q)}_{t-1}$ dependence in \eqref{eq:reg2}. Furthermore, we note that the variables $\bX_t, \bY_t, \bZ_t$ may depend on \emph{latent} (unknown) or \emph{exogenous} (unmeasured) variables not included in the regressions. Rather, we should view the regressions purely as \emph{predictive models} specified by the parameters $A$ and $A'$ respectively, which are then to be fitted by an OLS or equivalent procedure.}. The \emph{G-causality} of $\bY$ to $\bX$ given $\bZ$ is a measure of the extent to which inclusion of $\bY$ in the second model \eqref{eq:reg2} reduces the prediction error of the first model \eqref{eq:reg1}.

The standard measure of G-causality in the literature is defined for \emph{univariate} predictor and predictee variables $Y$ and $X$, and is given by the natural logarithm of the ratio of the residual variance in the restricted regression \eqref{eq:reg1} to that of the unrestricted regression \eqref{eq:reg2}. In our notation \footnote{Note that even though $X$ and $Y$ are univariate, the \emph{lagged} variables $\blX$ and $\blY$ will generally be multivariate (at least if $p,q>1$); hence they are written in bold type.}
\begin{eqnarray}
    \cgc Y X\bZ
	& \equiv & \lnt{\frac{\var{\eps_t}}{\var{\eps'_t}}} \nonumber \\
	& = & \lnt{\frac{\Covs{\eps_t}}{\Covs{\eps'_t}}} \nonumber \\
	& = & \lnt{\frac{\cCov X{\blX\mc\blZ}}{\cCov X{\blX\mc\blY\mc\blZ}}} \label{eq:gcu}
\end{eqnarray}
where the last equality follows from the general formula \eqref{eq:rescov}. By stationarity this expression does not depend on time $t$, so we drop the subscript when there is no danger of confusion. Note that the residual variance of the first regression will always be larger than or equal to that of the second, so that $\cgc Y X\bZ \ge 0$ always. As regards statistical inference, it is known that the corresponding maximum likelihood estimator $\ecgc Y X\bZ$ will have (asymptotically for large samples) a $\chi^2$-distribution under the null hypothesis  $\cgc Y X\bZ = 0$ \cite{Granger63,Whittle53} and a non-central  $\chi^2$-distribution under the alternative hypothesis $\cgc Y X\bZ > 0$ \cite{Geweke82,Wald43}.

Although rarely considered in the literature, there is no requirement in principle that either the predictee or predictor variable be univariate. In this Letter we address the general case where all variables are allowed to be \emph{multivariate}; see \cite{Feng09} and \cite{Barnett:2009b} for motivation and discussion regarding this generalization. For the case of a multivariate predictor, \eeqref{eq:gcu} above (with $Y$ replaced by the bold-type $\bY$) is a valid and consistent formula for G-causality. However, generalization to the case of a multivariate predictee is less clear cut and there does not yet appear to be a standard definition for G-causality in the literature. Here we use an extension first proposed by Geweke \cite{Geweke82}, in which the residual variance $\var{\eps_t} = \Covs{\eps_t}$ is replaced by the generalized variance $\dett{\Covs{\beps_t}}$:
\begin{eqnarray}
    \cgc\bY\bX\bZ
	& \equiv & \lnt{\frac{\dett{\Covs{\beps_t}}}{\dett{\Covs{\beps'_t}}}} \nonumber \\
	& = & \lnt{\frac{\dett{\cCov{\bX}{\blX\mc\blZ}}}{\dett{\cCov{\bX}{\blX\mc\blY\mc\blZ}}}} \label{eq:gc}
\end{eqnarray}
This formula always produces a non-negative quantity, and for a univariate predictee reduces to \eqref{eq:gcu}. Moreover, its estimator is also asymptotically $\chi^2$-distributed. Geweke \cite{Geweke82} lists a number of motivations for this choice, to which we add the result presented in this Letter. (An alternative formulation for multivariate G-causality is proposed in \cite{Feng09}, although see \cite{Barnett:2009b} for more detailed discussion and further motivation for the form \eqref{eq:gc}.)
\newline

With $\bX_t, \bY_t, \bZ_t$ as before, the  \emph{transfer entropy} of $\bY$ to $\bX$ given $\bZ$ \cite{schreiber00, KaiserSchreiber02}
is defined as the difference between the entropy of $\bX$ conditioned on its own past and the past of $\bZ$, and its entropy conditioned, in addition, on the past of $\bY$:
\begin{equation}
    \cxe\bY\bX\bZ \equiv \centro{\bX}{\blX\mc\blZ} - \centro{\bX}{\blX\mc\blY\mc\blZ} \label{eq:te}
\end{equation}
where $\entro\cdot$ denotes entropy and $\centro\cdot\cdot$ conditional entropy. Again, by stationarity transfer entropy does not depend on time $t$, and $\cxe\bY\bX\bZ \ge 0$ always. $\cxe\bY\bX\bZ$ may be understood as the degree of uncertainty of $\bX$ resolved by the past of $\bY$ over and above the degree of uncertainty of $\bX$ resolved by its \emph{own} past. As with Granger causality, the transfer entropy literature generally deals only with univariate variables, although in this case the extension \eqref{eq:te} to the multivariate case is unproblematic.
\newline

We now turn to the equivalence with G-causality. For a \emph{multivariate Gaussian} random variable $\bX$ we have the well-known expression \cite{PapoulisPillai02}
\begin{equation*}
    \entro\bX = \shalf \lnt{\dett{\Covs\bX}} + \shalf n\lnt{2\pi e}
\end{equation*}
for entropy in terms of the determinant of the covariance matrix, where $n$ is the dimension of $\bX$. We now show that the conditional entropy $\centro\bX\bY$ for two jointly multivariate Gaussian variables may be expressed in terms of the determinant of the corresponding partial covariance matrix:
\begin{equation}
    \centro\bX\bY = \shalf \lnt{\dett{\cCov\bX\bY}} + \shalf n\lnt{2\pi e} \label{eq:Gcent}
\end{equation}
To see this, we have
\begin{eqnarray*}
    \centro\bX\bY
    &\equiv& \entro{\bX\mc\bY} - \entro\bY \\
	&=& \shalf \lnt{\dett{\Covs{\bX\mc\bY}}} - \shalf \lnt{\dett{\Covs\bY}} \\
	&& + \shalf n\lnt{2\pi e}
\end{eqnarray*}
Now
\begin{equation*}
    \Covs{\bX\mc\bY} = \begin{pmatrix} \Covs\bX & \Cov\bX\bY \\ \trans{\Cov\bX\bY} & \Covs\bY \end{pmatrix}
\end{equation*}
and from the block determinant identity \cite{Horn85}
\begin{equation*}
    \dett{\begin{matrix} A & B \\ C & D \end{matrix}} = \dett D \dett{A-BD^{-1}C}
\end{equation*}
we have
\begin{equation*}
	\dett{\Covs{\bX\mc\bY}} = \dett{\Covs\bY} \cdot \dett{\cCov\bX\bY}
\end{equation*}
from which we obtain \eqref{eq:Gcent} \footnote{We note that for jointly multivariate Gaussian variables $\bX, \bY$ a standard result states that the covariance matrix of $\bX$ given $\bY = \by$ does not depend on the value of $\by$, so that the \emph{conditional covariance} $\ccovs\bX\bY$ is a well-defined (non-random) covariance matrix, which is just the partial covariance $\cCov\bX\bY$ of \eqref{eq:ccxy}.}.

If, then, the processes $\bX_t, \bY_t, \bZ_t$ are jointly multivariate Gaussian (\ie\ any finite subset of the component variables $X_{ti}, Y_{s\alpha}, Z_{ua}$ has a joint Gaussian distribution) it follows from \eqref{eq:Gcent} that the expression \eqref{eq:te} for transfer entropy becomes \footnote{This is essentially a multivariate, conditional version of the formula given in \cite{KaiserSchreiber02}, eq. (19).}
\begin{equation}
    \cxe\bY\bX\bZ \equiv \shalf \lnt{\frac{\dett{\cCov{\bX}{\blX\mc\blZ}}}{\dett{\cCov{\bX}{\blX\mc\blY\mc\blZ}}}} \label{eq:Gte}
\end{equation}
Comparing \eqref{eq:Gte} with \eqref{eq:gc} leads directly to our central result: if all processes are jointly Gaussian, then
\begin{equation}
    \cgc\bY\bX\bZ= 2\cxe\bY\bX\bZ \label{eq:tegceq}
\end{equation}
so that \emph{Granger causality and transfer entropy are equivalent up to a factor of $2$}. This result holds, in particular, for a univariate predictee $X$ with the standard definition \eqref{eq:gcu} of G-causality.
\newline

Empirically, numerical equivalence between G-causality and transfer entropy will depend on the method used to estimate the transfer entropy in sample. If it is assumed at the outset that the data may be reasonably modeled as Gaussian---and that, consequently, conditional entropies may be estimated from the appropriate sample covariance matrices---then, of course, numerical equivalence will be guaranteed. If, however, conditional entropies are estimated directly from sampled probability distributions, results will vary with the estimation technique. It is known that naive estimation of transfer entropy by partitioning of the state space is problematic \cite{schreiber00} and that such estimators frequently fail to converge to the correct result \cite{KaiserSchreiber02}. In practice, more sophisticated techniques such as kernel \cite{Silverman86} or $k$-nearest neighour estimators \cite{Kraskov04,Frenzel07}, will need to be deployed; however, such techniques may entail their own assumptions about the empirical distribution of the data (see \cite{KaiserSchreiber02} for a good discussion on these points). Furthermore, unlike G-causality, for which the (asymptotic) distribution of the sample statistic is known, we are not aware of any such general result for transfer entropy. Thus in particular significance testing for transfer entropy estimates is likely to be hard.
\newline

Our result \eqref{eq:tegceq} provides for the first time a unified framework for data-driven causal inference that bridges information-theoretic and autoregressive methods. In particular, it opens new research possibilities in transforming findings originally developed in one domain into the other. For example, an advantage of the autoregressive approach is that it admits a straightforward decomposition by frequency \cite{Geweke82,DingEtal06}. Our result now provides a foundation for the development of spectral implementations of transfer entropy. In the opposite direction, the invariance of information-theoretic quantities under general nonlinear transformations \cite{KaiserSchreiber02} could potentially prove useful in the identification of appropriate nonlinear autoregressive models \cite{ChenEtal04,AnconaEtal04}. Preliminary work by the authors indicates, perhaps surprisingly, that under Gaussian assumptions there is nothing extra to account for by nonlinear extensions to G-causality, since a stationary Gaussian AR process is necessarily linear \cite{Barnett:2009b}. This finding has practical significance because sensitivity to nonlinear data features is often presented as a reason to prefer transfer entropy to G-causality (see \eg\ \cite{SpornsLungarella06}).

As regards Gaussian assumptions, although their appropriateness may be disputed in the context of specific physical systems, they are nevertheless widely employed in neuroscience, econometrics and beyond, frequently in the role of an analytical benchmark for subsequent more physically motivated analysis. In practice, given empirical data it is likely to be difficult to establish the extent to which Gaussian assumptions are tenable, particularly for highly multivariate datasets and limited sample sizes. Further research is thus required to characterize---both analytically and in sample---the manner in which the equivalence \eqref{eq:tegceq} breaks down when Gaussian assumptions fail. As a starting point it is known, at least, that in the generic (non-Gaussian) case, nonzero G-causality implies nonzero transfer entropy \cite{Marinazzo08}.

More generally, G-causality is typically implemented within the well-understood and easily applicable framework of MVAR modeling. This implementation, however, implies many assumptions about how to model the data. Transfer entropy by contrast, although on a theoretical level ``model agnostic'' (in the sense that it involves no presumptions about the joint statistical distribution of the data), may present severe difficulties in empirical application. Investigators, then, are free to use whichever practical methods best suit their data. Numerical issues aside, the analytical equivalence \eqref{eq:tegceq} furnishes the essential point that---under Gaussian assumptions---G-causality has a natural interpretation as transfer entropy and \textit{vice-versa}.

\begin{acknowledgments}
\textbf{Acknowledgments.} AKS is supported by EPRSC Leadership Fellowship EP/G007543/1, which also supports the work of ABB.
\end{acknowledgments}

\bibliography{tegc_PRL}

\end{document}